\documentstyle[aps,preprint]{revtex}
\begin{document}
\draft
\hsize\textwidth\columnwidth\hsize\csname @twocolumnfalse\endcsname

\title{Does the Hubbard Model Show $d_{x^2-y^2}$ Superconductivity?}
\author{Gang Su$^{\ast}$ and Masuo Suzuki$^{\dag}$}
\address{ Department of Applied Physics, Faculty of Science,
 Science University of Tokyo\\
1-3, Kagurazaka, Shinjuku-ku, Tokyo 162, Japan}
\date{Received 9 October 1997}

\maketitle

\begin{abstract}
It is rigorously shown that the
two-dimensional Hubbard model with narrow bands (including next
nearest-neighbor hopping, etc.) does not exhibit
$d_{x^2 -y^2}$-wave pairing long-range order at any nonzero temperature.
This kind of pairing long-range order will also be excluded at zero
temperature if an excited energy gap opens
in the charge excitation spectrum of the system.
These results hold true for both
repulsive and attractive Coulomb interactions and for any electron fillings,
and are consistent with quantum Monto Carlo calculations.\\

\end{abstract}

\pacs{PACS numbers: 74.20.-z, 05.30.-d}


Considerable experimental evidence shows that the dominant
symmetry of the pairing order
parameter in high temperature superconductors may be $d_{x^2-y^2}$-wave
(see, e.g., Ref.\cite{shen} for a review). Though some
disputes\cite{ander}
regarding this challenging
issue still remain, a great number of people arrive at such a consensus
that exploring the
possibility of d-wave superconductivity in strongly correlated electrons
would be quite useful towards
ultimately successfully explaining high temperature superconductivity,
thus resulting in numerous studies on this subject. As a matter
of fact, owing to its apparent simplicity the two-dimensional (2D)
Hubbard model naturally
becomes an
actively debating focus in recent years, as it is widely thought to provide a
simple model to interpret some essential features relevant to the physical
properties of $CuO_2$ planes in the cuprate oxides. In spite of intense efforts
being made
both numerically and analytically\cite{scala1,dago}, however, a basic
question whether or
not the
$d_{x^2-y^2}$-wave
pairing long-range order (LRO) in the 2D Hubbard model
exists, is still
inconclusive. Actually, at energy scales and lattice sizes accessible
to numerical
simulations no definite sign of $d_{x^2-y^2}$ superconductivity has been
detected in
this model, while some analytical works using different approximations
appear to
suggest positive answers\cite{scala1,dago,scala2}, thereby leaving some
controversies
and ambiguities to be resolved. To clarify them,
 rigorous results are particularly needed at this stage.

In this paper, based on Bogoliubov's inequality we show rigorously that
the  2D Hubbard model with narrow bands (including next
nearest-neighbor
hopping, etc.) does not exhibit
$d_{x^2 -y^2}$-wave pairing LRO at any nonzero temperature.
This kind of pairing LRO will also be excluded if a gap opens
in the charge excitation spectrum of the system. These results
hold true for both repulsive and attractive Coulomb interactions
and for any electron fillings. Combining with other known exact results, one
would conclude that the 2D Hubbard model might not have the right stuff
for describing superconductivity in the cuprate oxides in this sense
provided that the superconducting mechanisms in these materials are
supposed to be
due to condensation of either s-wave Cooper, or generalized $\eta$ or
$d_{x^2-y^2}$-wave electron pairs. The present observations are consistent with
quantum Monto Carlo results.

Let us start with some preliminary definitions. The $d_{x^2 -y^2}$-wave pairing
operator (like the Cooper pairing operator) can be defined as\cite{note}
\begin{eqnarray}
\Delta_{d}^{+} = \sum_{\bf k}(\cos k_x - \cos k_y) c_{{\bf
k}\uparrow}^{\dagger}
c_{-{\bf k}\downarrow}^{\dagger} = \frac12 \sum_{{\bf r},\bbox{\bf
\delta}}
f(\bbox{\bf \delta})
c_{{\bf r}\uparrow}^{\dagger}c_{{\bf r}+\bbox{\bf
\delta}\downarrow}^{\dagger},
~~~\Delta_d^- = (\Delta_{d}^{+})^{\dagger},
\label{d-operator}
\end{eqnarray}
where we have used the Fourier transform of an electron operator $c_{{\bf
k}\sigma}$
\begin{eqnarray}
c_{{\bf k}\sigma} = \frac{1}{\sqrt{M}}\sum_{j} \exp (i{\bf k}\cdot
{\bf R}_j) c_{j\sigma}, ~~~
c_{j\sigma} = \frac{1}{\sqrt{M}}\sum_{\bf k} \exp (-i{\bf k}\cdot
{\bf R}_j) c_{{\bf k}\sigma},
\label{cfourier}
\end{eqnarray}
and $f(\bbox{{\bf \delta}}) = +1$ $(-1)$ for $\bbox{{\bf \delta}} = \pm {\bf
a}_x$ $(\pm {\bf a}_y)$, and zero otherwise, where ${\bf a}_x$, ${\bf a}_y$ are
unit vectors connecting nearest-neighbor sites, $M$ is the number of
lattice sites,
${\bf R}_j$ (${\bf r}$) is the position vector of $j$th ($r$th) site, and
$\sigma$ denotes spin. In the following we for simplicity take
$\bbox{{\bf \delta}} = \pm {\bf a}_x$, $\pm {\bf a}_y$.
According to Bogoliubov\cite{bogo}, when one studies a degenerate
state of the statistical equilibrium, one should first
remove the degeneracy by introducing a symmetry-breaking field, and then
turn to investigate the so-called quasi-averages involved.
On account of this reason we define the $d_{x^2-y^2}$-wave pairing order
parameter per site as
\begin{eqnarray}
g = \lim_{\nu \rightarrow 0^+}\lim_{M \rightarrow \infty} \langle
\frac{\Delta_d^+}{M}
\rangle,
\label{d-order}
\end{eqnarray}
where $\langle \cdots \rangle$ stands for the thermal average in a grand
canonical
ensemble, and $\nu$ is the amplitude of a U(1) symmetry-breaking field.
Note that
the two limit processes are non-interchangeable. The nonvanishing of $g$,
namely $g \neq 0$ means the
existence of $d_{x^2-y^2}$-wave pairing LRO, while $g=0$ gives the converse
result.

We now consider a general Hubbard model with narrow bands on a periodic
lattice in the presence of a U(1) symmetry-breaking field.
The Hamiltonian reads
\begin{eqnarray}
& &H = \sum_{i,j}\sum_{\sigma} T({\bf R}_i - {\bf R}_j)
c_{i\sigma}^{\dag}c_{j\sigma} + \sum_{i} U_i n_{i\uparrow} n_{i\downarrow}
-\sum_i \mu_i (n_{i\uparrow} + n_{i\downarrow}) \nonumber\\
& & \hspace{0.5cm} - \nu (\Delta_d^+ + \Delta_d^-),
\label{hamilt}
\end{eqnarray}
where the sum on $i$ and $j$ can run over all $M$ lattice sites,
$c_{i\sigma}^{\dag}$ ($c_{i\sigma}$) is the creation (annihilation)
operator for an electron at site $i$ with spin $\sigma$,
the on-site Coulomb interaction $U_i$ and the chemical potential $\mu_i$
are allowed to be position-dependent for generality, and $n_{i\sigma}=
c_{i\sigma}^{\dagger}c_{i\sigma}$, the number operator of electrons.
 $T({\bf R}_i - {\bf R}_j)$, which has a property $T({\bf R}_i - {\bf R}_j)
= T^{*}({\bf R}_j - {\bf R}_i)$, is the local
overlap integral which designates the energy bands of the model.
In fact we require that $T({\bf R}_i - {\bf R}_j)$ survives only for short-r
anged
overlapping in the present case. The last term in Eq.(\ref{hamilt}) introduces
the
effect of the U(1) symmetry-breaking field, where $\nu$ is an infinitesimal
quantity, and $\Delta_{d}^{\pm}$
are given by Eq.(\ref{d-operator}).

As we attempt to explore the possibility of
$d_{x^2-y^2}$ pairing LRO in Eq.(\ref{hamilt}),
in the following we shall use Bogoliubov's
inequality\cite{bogo1,mermin}
\begin{eqnarray}
|\langle [C^{\dag}, A^{\dag}]\rangle|^2 \leq \frac{\beta}{2}\langle \{A,
A^{\dag}\}\rangle \langle [[C, H], C^{\dag}]\rangle,
\label{bogoineq}
\end{eqnarray}
for any quantum-mechanical operators $A$ and $C$, and with
the inverse temperature $\beta = 1/T$ ($k_{B}=1$), where $[,]$ and
$\{,\}$ are the usual commutator and anticommutator, respectively.
This inequality
was used to exclude the possibility of magnetic LRO in the
Heisenberg \cite{mermin} and Hubbard \cite{walker,ghosh} models as well as
superfluidity in Fermi liquids \cite{hohen} in one and two dimensions at
nonzero temperature. Quite recently, this inequality was also used to
exclude the possibility of s-wave Cooper pairing and generalized $\eta$
pairing LRO in the 1D and 2D Hubbard models with narrow bands at nonzero
temperature\cite{su}. It should be mentioned that the SU(2)
Lie algebra obeyed by the relevant operators (e.g., spin operators,
Cooper pairing operators and $\eta$ pairing operators, etc.) plays a key
role in applying this inequality to the above-mentioned cases.
However, one may observe that the $d_{x^2-y^2}$-wave pairing operators
defined in Eq.(\ref{d-operator}) do not obey the SU(2) symmetry, which
makes Bogoliubov's inequality not directly applicable to the present case,
as stated in Ref.\cite{su}.
Fortunately, this difficulty can be overcome by noting the simple
fact that $f(\bbox{{\bf \delta}})$
in Eq.(\ref{d-operator}) takes values either $+1$ and $-1$ or zero so that
we can decompose the $d_{x^2-y^2}$ pairing order parameter per site, $g$,
into four terms each of which obeys the SU(2) algebra. It is this property
that makes it possible for applying inequality (\ref{bogoineq}) to our
case\cite{note2}. We would like to mention here that
although one
can find the standard derivations in Refs.\cite{mermin,walker,ghosh,su},
for reader's convenience and for this paper being self-contained,
we shall below intend to present our calculations in some detail.

We define the following operators
\begin{eqnarray}
\tilde{\eta}_{\bf r}^{+} = c_{{\bf r}\uparrow}^{\dag}c_{{\bf r}+\bbox{{\bf
\alpha}}
\downarrow}^{\dag}, ~~~\tilde{\eta}_{\bf r}^{-} =
c_{{\bf r}+\bbox{{\bf \alpha}}\downarrow}c_{{\bf r}\uparrow},
~~~\tilde{\eta}_{{\bf r}}^{z} = \frac12 (n_{{\bf r}\uparrow}
+ n_{{\bf r}+\bbox{{\bf \alpha}}\downarrow} -1),
\label{etaope}
\end{eqnarray}
with an arbitrary constant vector $\bbox{{\bf \alpha}}$ on the lattice. It
can be
verified that they satisfy
\begin{eqnarray}
[\tilde{\eta}_{\bf r}^{+}, \tilde{\eta}_{{\bf r}'}^{-}] = 2
\tilde{\eta}_{\bf r}^{z}
\delta_{{\bf r}{\bf r}'}, ~~~~[\tilde{\eta}_{\bf r}^{\pm},
\tilde{\eta}_{{\bf r}'}^{z}]
= \mp \tilde{\eta}_{\bf r}^{\pm}\delta_{{\bf r}{\bf r}'}.
\label{etacomm}
\end{eqnarray}
The Fourier transforms of $\tilde{\eta}$ operators, like spin operators in
Refs.\cite{mermin,ghosh}, are defined by
\begin{eqnarray}
\tilde{\eta}_{\bf r}^{\pm,z} = \frac{1}{M}\sum_{{\bf k}} \exp (-i {\bf
k}\cdot
{\bf r})\tilde{\eta}^{\pm,z}({\bf k}), ~~~~
\tilde{\eta}^{\pm,z}({\bf k}) =
\sum_{\bf r} \exp (i {\bf k}\cdot {\bf r})\tilde{\eta}_{\bf r}^{\pm,z},
\label{etafourier}
\end{eqnarray}
which comply
\begin{eqnarray}
[\tilde{\eta}^{+}({\bf k}), \tilde{\eta}^{-}({\bf k}')]
= 2 \tilde{\eta}^{z}({\bf k}+{\bf k}'), ~~~
[\tilde{\eta}^{\pm}({\bf k}), \tilde{\eta}^{z}({\bf k}')]
= \mp \tilde{\eta}^{\pm}({\bf k}+{\bf k}').
\label{etafourcomm}
\end{eqnarray}
Eqs.(\ref{etafourcomm}) come from Eqs.(\ref{etacomm}) and
(\ref{etafourier}).

With these definitions we choose $A = \tilde{\eta}^{+}(-{\bf k}-{\bf Q})$
and
$C = \tilde{\eta}^{z}({\bf k})$ in (\ref{bogoineq}) for our purposes.
After some algebra
for the double-commutator one gets the inequality,
\begin{eqnarray}
\langle [[\tilde{\eta}^{z}({\bf k}), H], \tilde{\eta}^{z}(-{\bf
k})]\rangle &\leq&
    \frac12 \sum_i |T({\bf R}_i)| |\cos({\bf k} \cdot {\bf R}_i)-1|
    |\sum_{{\bf k}'\sigma} e^{i{\bf k}'\cdot {\bf R}_i} \langle
    c_{{\bf k}' \sigma}^{\dag}c_{{\bf k}' \sigma}\rangle| \nonumber \\
& & \hspace{0.5cm} + |\nu| |\langle \Delta_d^{+} + \Delta_d^{-}\rangle|
\nonumber \\
&\leq & \frac{N}{4}\sum_i |T({\bf R}_i)|  R_i^2 k^2 + 2 |\nu| \cdot
|\langle \Delta_d^{+} \rangle|.
\label{etaineq1}
\end{eqnarray}
In the derivation of this inequality we have used the property of
translation
invariance and such a few simple facts as $ \sum_{{\bf k}'\sigma} \langle
c_{{\bf k}' \sigma}^{\dag}c_{{\bf k}' \sigma}\rangle = N $,
the total number of electrons, $1-\cos x < x^2/2$ and $\langle
\Delta_d^{+} \rangle
= \langle \Delta_d^{-} \rangle^{\dag}$. Substituting inequality
(\ref{etaineq1})
into (\ref{bogoineq}) we have
\begin{eqnarray}
\frac{1}{M} \langle \{\tilde{\eta}^{+}(-{\bf Q}-{\bf k}),
\tilde{\eta}^{-}({\bf Q}+{\bf k})\} \rangle \geq
\frac{2 |{\cal F}_{\nu,M}({\bf Q}, \bbox{{\bf \alpha}})|^2}{\beta ({\xi}
k^2 + 2 |\nu|
|\langle \Delta_d^{+}/M \rangle|)},
\label{etacommineq}
\end{eqnarray}
where ${\cal F}_{\nu,M}({\bf Q}, \bbox{{\bf \alpha}}) = \langle
\tilde{\eta}^{+}
({\bf Q}) \rangle /M$, and
$\xi =(N/M) \sum_{i}|T({\bf R}_i)|{\bf R}_i^2 /4$.
Since the
$T({\bf R}_i)$'s are the matrix elements of the overlap integral between
Wannier functions which decrease rapidly with distance for strongly
correlated
electrons, the summation $\sum_{i}|T({\bf R}_i)|{\bf R}_i^2$ is well defined.
For the single-band Hubbard model as well as one with next nearest-neighbor
hopping integral the values of $\sum_{i}|T({\bf R}_i)|{\bf R}_i^2$ on
a hypercubic lattice can be found in Ref.\cite{su}.
Summing both sides of the above inequality over ${\bf k}$
and noting that
$(1/M)\sum_{{\bf k}}\langle \{A,A^\dagger\} \rangle =\sum_{{\bf r}}
\langle
\{\tilde{\eta}_{{\bf r}}^{+},\tilde{\eta}_{{\bf r}}^{-} \} \rangle
= \sum_{{\bf r}} \langle [1 - (n_{{\bf r}\uparrow}-n_{{\bf r}+
\bbox{{\bf \alpha}}\downarrow})^2]\rangle \leq M$,
we obtain
\begin{eqnarray}
|{\cal F}_{\nu,M}({\bf Q}, \bbox{{\bf \alpha}}) |^2
\leq \frac{\beta M}{2} \left( \sum_{{\bf k}} \frac{1}{
\xi k^2 + 2|\nu| |\langle \Delta_d^{+}/M \rangle|} \right)^{-1}.
\label{fineq}
\end{eqnarray}
Now we take the thermodynamic limit, i.e., $N \rightarrow \infty$ and
$M \rightarrow \infty$ with the ratio $N/M$ fixed.
Then the sum on ${\bf k}$ in (\ref{fineq}) can be replaced
by the integral over the first Brillouin zone. Suppose that $k_0$ is the
distance
of the nearest Bragg plane from the origin in ${\bf k}$ space. Then
we obtain for small $\nu$ the following inequalities
\begin{eqnarray}
|{\cal F}_{\nu,\infty}({\bf Q}, \bbox{{\bf \alpha}})| &\leq&
(\xi \beta^2)^{1/4} |\nu|^{1/4}, ~~~~~~~~~(1D)
\label{ineq1d}\\
 |{\cal F}_{\nu,\infty}({\bf Q}, \bbox{{\bf \alpha}})| &\leq&
\sqrt{\frac{\xi \beta}{\pi}}
\frac{1}{|\ln |\nu||^{1/2}}, ~~~~~~~~(2D) \label{ineq2d}
\end{eqnarray}
where we have used $\lim_{M \rightarrow \infty} |\langle \Delta_d^{+}/M
\rangle|
\leq 2$ in (\ref{ineq1d}). Inequalities (\ref{ineq1d}) and (\ref{ineq2d})
tell us that $|{\cal F}_{0,\infty}({\bf Q}, \bbox{{\bf \alpha}})| =0$ for
any nonzero temperature as the U(1) symmetry-breaking field is turned off
($\nu \rightarrow 0^+$). (Recall that the thermodynamic limit has been
taken before we remove the U(1) symmetry-breaking field.)
This means that
${\cal F}_{0,\infty}({\bf Q}, \bbox{{\bf \alpha}}) := \lim_{\nu \rightarrow
0^+}
\lim_{M \rightarrow \infty} \langle (1/M) \sum_{{\bf r}}\exp {(i {\bf Q}
\cdot
{\bf r})}c_{{\bf r}\uparrow}^{\dag}c_{{\bf r}+\bbox{{\bf
\alpha}}\downarrow}^{\dag}
\rangle_{\nu,M} =0$ for {\it any} possible ${\bf Q}$ and $\bbox{{\bf \alpha}}$
for $T>0$ in the
1D and 2D Hubbard models defined in Eq.(\ref{hamilt}). On the other hand,
we note that the $d_{x^2-y^2}$-wave pairing order parameter per site can be
rewritten
as $g = (1/2) \sum_{\bbox{{\bf \delta}}}f(\bbox{{\bf \delta}})
{\cal F}_{0,\infty}({\bf 0}, \bbox{{\bf \delta}})$ because the existence of
${\cal F}_{0,\infty}({\bf 0}, \bbox{{\bf \delta}})$ has been proved, where
we have set ${\bf Q} = {\bf 0}$ and $\bbox{{\bf \alpha}} = \bbox{{\bf
\delta}}$. Here use has been made of the well-known theorem: $\lim a \pm \lim b
= \lim (a\pm b)$ if $\lim a$ and $\lim b$ exist.
Since ${\cal F}_{0,\infty}({\bf 0}, \bbox{{\bf \delta}})=0$ for $T>0$
and $f(\bbox{{\bf \delta}})$,
by virtue of its definition, takes either $\pm 1$ or zero, we finally have
$g=0$ for $T>0$. Consequently, we have proved that the 2D Hubbard model with
narrow
bands, defined by Eq.(\ref{hamilt}), does not show $d_{x^2-y^2}$-wave pairing
LRO
at any nonzero temperature.

When $\beta/2$ is replaced by $1/E_{gap}$ in inequality (\ref{bogoineq}),
where
$E_{gap}$ $(>0)$ is an excitation energy gap between the lowest excited
state and the ground state of the system,
(\ref{bogoineq})
still holds true at zero temperature. We refer to Refs.\cite{su,assa} for
detail discussions. Therefore, all the above analyses can apply to the case at
zero temperature except $\beta/2$ replaced by $1/E_{gap}$. Then we could
conclude
that if an energy gap opens in the charge excitation spectrum of the
system
(\ref{hamilt}), there will also be no $d_{x^2-y^2}$-wave pairing LRO in two
dimensions at zero temperature.

A few remarks are in order. (i) The derivations above presented are valid
for
both repulsive and attractive on-site Coulomb interactions
 and for any electron filling fraction.
(ii) The method used in this paper can be readily extended to exclude
the possibility of extended s-wave [with pairing operator
$\Delta_{e-s}^{+} = \sum_{\bf k}(\cos k_x + \cos k_y) c_{{\bf
k}\uparrow}^{\dagger}
c_{-{\bf k}\downarrow}^{\dagger}$] superconductivity
in the 1D and 2D Hubbard models at finite temperatures. In the Hilbert subspace
without
doubly-occupied sites this method
might be applied to the t-J model as well. It should be emphazied
that the method works only for low dimensions (1D and 2D), not for three
dimensions.
(iii) The conclusions drawn in this
paper are quite consistent with the results from numerical simulations
(e.g., quantum Monte Carlo calculations)
in the 2D Hubbard model. We refer to
Refs.\cite{scala1,dago,scala2} for excellent reviews. The practical
situation is
that although most  numerical works show some
tendencies favouring $d_{x^2 -y^2}$ superconductivity,
but no definite sign of LRO has been detected, as pointed
out in Refs.\cite{assaad,kuroki}. We would like to mention here that the
present conclusions are not incompatible with the quantum Monte Carlo
observations that the long-tailed enhancements in the $d_{x^2-y^2}$
pairing correlation near half-filling\cite{kuroki} or the exhibition
of $d_{x^2-y^2}$-like pairing fluctuations at low temperatures\cite{scala3}
are detected in the 2D Hubbard model.
Recent analytic and quantum Monte Carlo
results
also show that the 2D Hubbard model with next
nearest-neighbor hopping integral does not exhibit any definite
sign of s-wave  and d-wave
superconductivity\cite{laca,veil,huss,zhang}, consistent with the present exact
result.
(iv) The nonexistence of s-wave Cooper pairing and generalized $\eta$
pairing LRO
at finite temperatures in the 1D and
2D Hubbard models with narrow bands has been proved in Ref.\cite{su,koma}.
Combining
these exact results one would conclude that the 2D Hubbard model might not
have the
right stuff for explaining high temperature superconductivity in the layered
cuprate oxides if the superconducting mechanisms in these materials are supposed
to be due to
condensation of one of the above-mentioned electron pairs, as a successful t
heory
should describe unifyingly the properties
not only at zero temperature but also at finite temperatures. (v) To choose
a proper model which could exhibit $d_{x^2-y^2}$-wave pairing LRO in 2D, one
may consider
those which contain electron interactions not commuting with the operator
$\sum_{{\bf r}}
\exp({i {\bf k} \cdot {\bf r}}) n_{{\bf r}}$, like one investigated in
Ref.\cite{assaad}, because in this way  Bogoliubov's inequality becomes
ineffective.
Another possibility is that the coupling between layers in the cuprate oxides
might ought
to be considered, which could also make Bogoliubov's inequality ineffective.

To summarize, we show rigorously that, by means of Bogoliubov's
inequality,
the 2D Hubbard model with narrow bands
(including next nearest-neighbor hopping, etc.) does not exhibit
$d_{x^2 -y^2}$ wave pairing LRO at any nonzero temperature.
This kind of pairing LRO will also be excluded if an excited energy gap
opens in the charge excitation spectrum of the system. These results
are valid for both repulsive and attractive Coulomb interactions and for
any electron fillings.
Combining with known exact results obtained
previously one would conclude that the 2D Hubbard model might not have enough
right
stuff for describing superconductivity in the cuprate oxides if the
superconducting mechanisms in these materials are supposed to be due to
condensation of one of the aforementioned electron pairs.
The present observations
are consistent with quantum Monto Carlo results.

\acknowledgments

One of authors (GS) is grateful to the Department of Applied Physics,
Science University of Tokyo, for the warm hospitality, and to the Nishina
foundation for support. This work has been supported by the CREST
(Core Research for Evolutional Science and Technology) of the Japan
Science and Technology Corporation (JST).


\begin{references}
\bibitem[\mbox{$$}]{Permanent address.} $^\ast$On leave from Graduate School,
Chinese Academy of Sciences, Beijing, China.
Electronic address: gsu@ap.kagu.sut.ac.jp\\
$^{\dag}$Electronic address: msuzuki@ap.kagu.sut.ac.jp
\bibitem{shen}Z.-X. Shen and D.S. Dessau, Phys. Rep. {\bf 253}, 1 (1995)
and references therein.
\bibitem{ander}P.W. Anderson, D. Pines, D.J. Scalapino and Z.X. Shen,
Phys.
Today {\bf 47}, No.2, 11 (1994).
\bibitem{scala1}D.J. Scalapino, Phys. Rep. {\bf 250}, 329 (1995);
D. Van Harlingen, Rev. Mod. Phys. {\bf 67}, 515 (1995); and
references therein.
\bibitem{dago}E. Dagotto, Rev. Mod. Phys. {\bf 66}, 763 (1994) and
references therein.
\bibitem{scala2}D.J. Scalapino, in {\it Does the Hubbard model have the
right stuff?},
edited by R.A. Broglia and J.R. Schrieffer, Proceedings of the
International
School of Physics, Enrico Fermi, Course CXXI (North-Holland, Amsterdam,
1994).
\bibitem{note}Note that without loss of generality we adopt this simple
form for the $d_{x^2-y^2}$ pairing operator (see, e.g. Ref.\cite{dago}).
 If a more symmetrical
form, like one exploited in Refs.\cite{scala1,scala2}, is used, the present
method also works, and the
conclusion drawn in this paper still remains true.
\bibitem{bogo}N.N. Bogoliubov, Physica {\bf 26}, S1 (1960).
\bibitem{bogo1}N.N. Bogoliubov, Phys. Abh. Sowjetunion {\bf 6}, 1, 113,
229
(1962); see
also H. Wagner, Z. Phys. {\bf 195}, 273 (1966).
\bibitem{mermin}N.D. Mermin and H. Wagner, Phys. Rev. Lett. {\bf 17}, 1133
(1966).
\bibitem{walker}M.B. Walker and Th.W. Ruijgrok, Phys. Rev. {\bf 171},
513(1968).
\bibitem{ghosh}D.K. Ghosh, Phys. Rev. Lett. {\bf 27}, 1584 (1971); G.S.
Uhrig,
Phys. Rev. B {\bf 45}, 4738(1992).
\bibitem{hohen}P.C. Hohenberg, Phys. Rev. {\bf 158}, 383 (1967).
\bibitem{su}G. Su, A. Schadschneider and J. Zittartz, Phys. Lett. A{\bf
230},
99 (1997).
\bibitem{note2}This simple fact was unfortunately overlooked in
Ref.\cite{su}.
\bibitem{assa}Assa Auerbach, {\it Interacting Electrons and Quantum
Magnetism}
(Springer-Verlag, New York, 1994).
\bibitem{assaad}F.F. Assaad, M. Imada and D.J. Scalapino,
Phys. Rev. Lett. {\bf 77}, 4592 (1996).
\bibitem{kuroki}K. Kuroki and H. Aoki, preprint cond-mat/9707277.
\bibitem{scala3}D.J. Scalapino, preprints cond-mat/9606217, 9606218.
\bibitem{laca}R. Lacaze, A. Morel, B. Petersson and J. Schr\"oper, preprint
cond-mat/9610193.
\bibitem{veil}A.F. Veilleux, A. D\'are, L. Chen, Y. Vilk and A. Tremblay,
Phys. Rev. B {\bf 52}, 16255 (1995).
\bibitem{huss}T. Husslein, I. Morgenstern, D.M. Newns, P.C. Pattnaik,
J.M. Singer, and H.G. Matuttis, Phys. Rev. B {\bf 54}, 16179 (1996).
\bibitem{zhang}S. Zhang, J. Carlson and J.E. Gubernatis, Phye. Rev. Lett.
{\bf 78}, 4486(1997); Phys. Rev. B55, 7464 (1997).
\bibitem{koma}T. Koma and H. Tasaki, Phys. Rev. Lett. {\bf 68},
2348 (1992). It can be noted that the method presented in this paper
could be extended to cover the d-wave pairing case if a
similar argument to ours (i.e., dividing the d-wave pairing operators into four
terms) is taken into account. Our presentation, however, is simpler and
accessible to general reader.
\end{references}
\end{document}